\documentclass[11pt]{article}

\begin{document}
\thispagestyle{empty}
\newcommand{\be}{\begin{equation}}
\newcommand{\ee}{\end{equation}}
\newcommand{\sect}[1]{\setcounter{equation}{0}\section{#1}}
\newcommand{\vs}[1]{\rule[- #1 mm]{0mm}{#1 mm}}
\newcommand{\hs}[1]{\hspace{#1mm}}
\newcommand{\mb}[1]{\hs{5}\mbox{#1}\hs{5}}
\newcommand{\bea}{\begin{eqnarray}}
\newcommand{\eea}{\end{eqnarray}}
\newcommand{\wt}[1]{\widetilde{#1}}
\newcommand{\ux}[1]{\underline{#1}}
\newcommand{\ov}[1]{\overline{#1}}
\newcommand{\sm}[2]{\frac{\mbox{\footnotesize #1}\vs{-2}}
           {\vs{-2}\mbox{\footnotesize #2}}}
\newcommand{\prt}{\partial}
\newcommand{\eps}{\epsilon}\newcommand{\p}[1]{(\ref{#1})}
\newcommand{\R}{\mbox{\rule{0.2mm}{2.8mm}\hspace{-1.5mm} R}}
\newcommand{\Z}{Z\hspace{-2mm}Z}
\newcommand{\cd}{{\cal D}}
\newcommand{\cg}{{\cal G}}
\newcommand{\ck}{{\cal K}}
\newcommand{\cw}{{\cal W}}
\newcommand{\vj}{\vec{J}}
\newcommand{\vl}{\vec{\lambda}}
\newcommand{\vz}{\vec{\sigma}}
\newcommand{\vt}{\vec{\tau}}
\newcommand{\poiss}{\stackrel{\otimes}{,}}
\newcommand{\tx}{\theta_{12}}
\newcommand{\tb}{\overline{\theta}_{12}}
\newcommand{\zw}{{1\over z_{12}}}
\newcommand{\sqp}{{(1 + i\sqrt{3})\over 2}}
\newcommand{\sqm}{{(1 - i\sqrt{3})\over 2}}
\newcommand{\NP}[1]{Nucl.\ Phys.\ {\bf #1}}
\newcommand{\PLB}[1]{Phys.\ Lett.\ {B \bf #1}}
\newcommand{\PLA}[1]{Phys.\ Lett.\ {A \bf #1}}
\newcommand{\NC}[1]{Nuovo Cimento {\bf #1}}
\newcommand{\CMP}[1]{Commun.\ Math.\ Phys.\ {\bf #1}}
\newcommand{\PR}[1]{Phys.\ Rev.\ {\bf #1}}
\newcommand{\PRL}[1]{Phys.\ Rev.\ Lett.\ {\bf #1}}
\newcommand{\MPL}[1]{Mod.\ Phys.\ Lett.\ {\bf #1}}
\newcommand{\BLMS}[1]{Bull.\ London Math.\ Soc.\ {\bf #1}}
\newcommand{\IJMP}[1]{Int.\ J.\ Mod.\ Phys.\ {\bf #1}}
\newcommand{\JMP}[1]{Jour.\ Math.\ Phys.\ {\bf #1}}
\newcommand{\LMP}[1]{Lett.\ Math.\ Phys.\ {\bf #1}}
\newpage
\setcounter{page}{0} \pagestyle{empty} \vs{12}

\begin{center}
{\LARGE {\bf On the Octonionic $M$-algebra and superconformal
$M$-algebra.}}\\ {\quad}\\

\vs{10} {\large F. Toppan} ~\\ \quad
\\
 {\large{\em CBPF - CCP}}\\{\em Rua Dr. Xavier Sigaud
150, cep 22290-180 Rio de Janeiro (RJ)}\\{\em Brazil}\\ E-mail:
toppan@cbpf.br

\end{center}
{\quad}\\ \centerline{ {\bf Abstract}}

\vs{6}

It is shown that the $M$-algebra related with the $M$ theory comes
in two variants. Besides the standard $M$ algebra based on the
real structure, an alternative octonionic formulation can be
consistently introduced. This second variant has striking
features. It involves only $52$ real bosonic generators instead of
$528$ of the standard $M$ algebra and moreover presents a novel
and surprising feature, its octonionic $M5$ (super-$5$-brane)
sector is no longer independent, but coincides with the octonionic
$M1$ and $M2$ sectors. This is in consequence of the
non-associativity of the octonions. An octonionic version of the
superconformal $M$-algebra also exists.  It is given by
$OSp(1,8|{\bf O})$ and admits $239$ bosonic and $64$ fermionic
generators. It is speculated that the octonionic $M$-algebra can
be related to the exceptional Lie and Jordan algebras that
apparently play a special role in the Theory Of Everything.

\section{Introduction}    
The generalized supersymmetries going beyond the standard H{\L}S
scheme \cite{hls} admit the presence of bosonic abelian tensorial
central charges associated with the dynamics of extended objects
(branes). It is widely known since the work of \cite{kt} that
supersymmetries are related to division algebras. Indeed, even for
the generalized supersymmetries, classification schemes based on
the associative division algebras (${\bf R}$, ${\bf C}$, ${\bf
H}$) are now available, see \cite{fer}. For what concerns the
remaining division algebra, the octonions, much less is known due
to the complications arising from their non-associativity.
Octonionic structures were, nevertheless, investigated in
\cite{{fm},{cs}} in application to superstring theory.\par
Octonions are not just a curiosity. They are the maximal division
algebra. This fact alone already justifies that they should
receive the same kind of attention paid to, let's say, the maximal
supergravity. However, their importance is more than that, they
are at the very heart of many exceptional structures in
mathematics and can be held responsible for their existence. Among
these exceptional structures we can cite the $5$ exceptional Lie
algebras and the exceptional Jordan algebras. Indeed, the $G_2$
Lie algebra is the automorphism group of the octonions, while
$F_4$ is the automorphism group of the $3\times 3$
octonionic-valued hermitian matrices realizing the exceptional
$J_3({\bf O})$ Jordan algebra. $F_4$ and the remaining exceptional
Lie algebras $E_6$, $E_7$, $E_8$ are recovered from the so-called
``magic square Tits' construction" which associates a Lie algebra
to any given pair of division algebras, if at least one of these
algebras coincides with the octonionic algebra \cite{bs}.
\par
It has been pointed out several times, \cite{{wit},{ram}} that the
exceptional Lie algebras fit well into the grand-unification
scenario. Moreover, the $E_8$ Lie algebra enters, through the
$E_8\times E_8$ tensor product, the anomaly-free heterotic string,
while the $G_2$ holonomy of seven-dimensional manifolds is
required, on phenomenological basis, to produce $4$-dimensional
$N=1$ supersymmetric field theories by compactification of the
eleven dimensions. This partial list of scattered pieces of
evidence has brought to suggest, see e.g. \cite{ram}, that for
some deep reasons, Nature seems to prefer exceptional structures.
In this context it deserves to be mentioned the special role of
the exceptional Jordan algebra $J_3({\bf O})$, not only associated
to the unique consistent quantum mechanical system (in the Jordan
framework, see \cite{cmp}) based on a non-associative algebra, but
also leading to a unique matrix Chern-Simon theory of Jordan type,
see \cite{smo}. \par In this talk I will discuss the
investigations presented in \cite{{lt1},{lt2}} concerning the
possibility of realizing general supersymmetries in terms of the
non-associative division algebra of the octonions. In particular
in \cite{lt1} it was shown that the $M$ algebra which supposedly
underlines the $M$-theory comes in two (and only two, due to the
absence of the complex and of the quaternionic structures)
variants. Besides the standard realization of the $M$-algebra
which involves real spinors and makes therefore use of the real
structure, an alternative formulation, requiring the introduction
of the octonionic structure, is also possible and can be
exploited. This is made possible due to the existence of an
octonionic description for the Clifford algebra defining the
$11$-dimensional Minkowskian spacetime and its related spinors.
The features of this second variant, the octonionic
$M$-superalgebra, are puzzling. While it is not at all surprising
that it contains fewer bosonic generators, $52$, w.r.t. the $528$
of the standard $M$-algebra (this is after all expected, since the
imposition of an extra structure always puts a constraint on a
theory), what really came as an unexpected surprise is the fact
that new conditions, not present in the standard $M$-theory, are
now found. These conditions imply that the different brane-sectors
are no longer independent. The octonionic $5$-brane alone contains
the whole set of degrees of freedom and is therefore equivalent to
the octonionic $M1$ and $M2$ sectors. We can write this
equivalence, symbolically, as $M5\equiv M1+M2$. This result is
indeed very intriguing. It implies that quite non-trivial
structures are found when investigating the octonionic
construction of the $M$-theory. It is quite tempting to think that
the exceptional structures that we mentioned before should be
better understood from this octonionic variant of the $M$-algebra,
rather than the standard real $M$-algebra.\par The next passage
consists in defining the closed algebraic structure which realizes
the octonionic superconformal $M$-algebra. It turns out that the
$OSp(1,64)$ superconformal algebra of the real $M$-theory is
replaced in the octonionic case by the $OSp(1,8|{\bf O})$
superalgebra of supermatrices with octonionic-valued entries and
total number of $7+232=239$ bosonic generators.

\section{Octonionic Clifford algebras and spinors.}

In the $D=11$ Minkowskian spacetime, where the $M$-theory should
be found, the spinors are real and have $32$ components. Since the
most general symmetric $32\times 32$ matrix admits $528$
components, one can easily prove that the most general
supersymmetry algebra in $D=11$ can be presented as
\begin{equation}
\{Q_a,Q_b\}= (C\Gamma_\mu )_{ab}P^\mu +(C\Gamma_{[\mu\nu]})_{ab}
Z^{[\mu\nu]} +
(C\Gamma_{[\mu_1\dots\mu_5]})_{ab}Z^{[\mu_1\dots\mu_5]}
\label{Malg}
\end{equation}
(where $C$ is the charge conjugation matrix), while $Z^{[\mu\nu]}$
and $Z^{[\mu_1\dots\mu_5]}$ are totally antisymmetric tensorial
central charges, of rank $2$ and $5$ respectively, which
correspond to extended objects \cite{{bst},{dk}}, the $p$-branes.
Please notice that the total number of $528$ is obtained in the
r.h.s as the sum of the three distinct sectors, i.e.
\begin{eqnarray}
528&=& 11+66+462.\label{sectors} \end{eqnarray}
 The algebra
(\ref{Malg}) is called the $M$-algebra. It provides the
generalization of the ordinary supersymmetry algebra, recovered by
setting $Z^{[\mu\nu]} \equiv Z^{[\mu_1\dots\mu_5]}\equiv 0$.
\par
In the next section we will prove the existence of an octonionic
version of Eq.~(1).  For this purpose we need at first to
introduce the octonionic realizations of Clifford algebras and
spinors. They exist only in a restricted class of spacetime
signatures which includes the Minkowskian $(10,1)$ spacetime.\par
The most convenient way to construct realizations of Clifford
algebras is to iteratively derive them with the help of the
following algorithm, allowing the recursive construction of $D+2$
spacetime dimensional Clifford algebras by assuming known a $D$
dimensional representation. Indeed, it is a simple exercise to
verify that if $\gamma_i$'s denotes the $d$-dimensional Gamma
matrices of a $D=p+q$ spacetime with $(p,q)$ signature (namely,
providing a representation for the $C(p,q)$ Clifford algebra) then
$2d$-dimensional $D+2$ Gamma matrices (denoted as $\Gamma_j$) of a
$D+2$ spacetime are produced according to either
\begin{eqnarray}
 \Gamma_j &\equiv& \left(
\begin{array}{cc}
  0& \gamma_i \\
  \gamma_i & 0
\end{array}\right), \quad \left( \begin{array}{cc}
  0 & {\bf 1}_d \\
  -{\bf 1}_d & 0
\end{array}\right),\quad \left( \begin{array}{cc}
  {\bf 1}_d & 0\\
  0 & -{\bf 1}_d
\end{array}\right)\nonumber\\
&&\nonumber\\ (p,q)&\mapsto&
 (p+1,q+1).\label{one}
\end{eqnarray}
or
\begin{eqnarray}
 \Gamma_j &\equiv& \left(
\begin{array}{cc}
  0& \gamma_i \\
  -\gamma_i & 0
\end{array}\right), \quad \left( \begin{array}{cc}
  0 & {\bf 1}_d \\
  {\bf 1}_d & 0
\end{array}\right),\quad \left( \begin{array}{cc}
  {\bf 1}_d & 0\\
  0 & -{\bf 1}_d
\end{array}\right)\nonumber\\
&&\nonumber\\ (p,q)&\mapsto&
 (q+2,p).\label{two}
\end{eqnarray}
Some remarks are in order. The two-dimensional real-valued Pauli
matrices $\tau_A$, $\tau_1$, $\tau_2$ which realize the Clifford
algebra $C(2,1)$ are obtained by applying either (\ref{one}) or
(\ref{two}) to the number $1$, i.e. the one-dimensional
realization of $C(1,0)$. We have indeed
\begin{eqnarray}
&\tau_A= \left(\begin{array}{cc}0 &1\\ -1&0  \end{array}\right),
\quad \tau_1= \left(\begin{array}{cc}0 &1\\ 1&0
\end{array}\right), \quad
\tau_2= \left(\begin{array}{cc}1 &0\\ 0&-1  \end{array}\right).
\quad &\label{Pauli}
\end{eqnarray}
The above algorithms can be applied to ``lift" the Clifford
algebra $C(0,7)$, furnishing higher-dimensional Clifford algebras.
$C(10,1)$ is constructed by successively applying (no matter in
which order) (\ref{one}) and (\ref{two}) to $C(0,7)$. For what
concerns $C(0,7)$, it must be previously known. Two inequivalent
realizations of $C(0,7)$ can be constructed. The first one is
associative and admits a matrix realization. Without loss of
generality (the associative irreducible representation of $C(0,7)$
is unique) we can choose expressing it through
\begin{eqnarray}
C(0,7) &\equiv& \begin{array}{c}
  \tau_A\otimes\tau_1\otimes{\bf 1}_2, \\
  \tau_A\otimes\tau_2\otimes{\bf 1}_2, \\
  {\bf 1}_2\otimes \tau_A\otimes \tau_1,\\
  {\bf 1}_2\otimes \tau_A\otimes \tau_2,\\
  \tau_1\otimes{\bf 1}_2\otimes\tau_A,\\
  \tau_2\otimes{\bf 1}_2\otimes\tau_A,\\
  \tau_A\otimes\tau_A\otimes\tau_A.
\end{array}\label{c07}
\end{eqnarray}
On the other hand another, inequivalent, realization is at
disposal. It is based on the identification of the $C(0,7)$
Clifford algebra generators with the seven imaginary octonions
$\tau_i$ satisfying the algebraic relation
\begin{eqnarray}
\tau_i\cdot \tau_j &=& -\delta_{ij} + C_{ijk} \tau_{k},
\label{octonrel}
\end{eqnarray}
for $i,j,k = 1,\cdots,7$ and $C_{ijk}$ the totally antisymmetric
octonionic structure constants given by
\begin{eqnarray}
&C_{123}=C_{147}=C_{165}=C_{246}=C_{257}=C_{354}=C_{367}=1&
\end{eqnarray}
and vanishing otherwise. This octonionic realization of the
seven-dimensional Euclidean Clifford algebra will be denoted as
$C_{\bf O}(0,7)$. Similarly, the octonionic realization $C_{\bf
O}(10,1)$, obtained through the lifting procedure, is realized in
terms of $4\times 4$ matrices with octonionic entries.
\par
One should be aware of the properties of the non-associative
realizations of Clifford algebras. In the octonionic case the
commutators $\Sigma_{\mu\nu} =\relax [\Gamma_\mu, \Gamma_\nu]$ are
no longer the generators of the Lorentz group. They correspond
instead to the generators of the coset $SO(p,q)/G_2$, being $G_2$
the $14$-dimensional exceptional Lie algebra of automorphisms of
the octonions. As an example, in the Euclidean $7$-dimensional
case, these commutators give rise to $7=21-14$ generators,
isomorphic to the imaginary octonions. Indeed
\begin{eqnarray}
\relax [\tau_i,\tau_j]& = &2C_{ijk}\tau_k .\label{octcomm}
\end{eqnarray}
The algebra (\ref{octcomm}) is not a Lie algebra, but a Malcev
algebra (due to the alternativity property satisfied by the
octonions, a weaker condition w.r.t. associativity, see
\cite{gk}). It can be regarded \cite{{hl},{cp}} as the ``quasi"
Lorentz algebra of homogeneous transformations acting on the seven
sphere $S^7$. This is so because $S^7$ is a parallelizable
manifold with a quasi (due to the lack of associativity) group
structure which can be identified with the unit octonions
\begin{eqnarray}
X^\dagger\cdot X&=&1.\label{uninorm}
\end{eqnarray}
Here $X^\dagger$ denotes the principal conjugation for the
octonions, namely
\begin{eqnarray}
X &=& x_0 + x_i\tau_i,\nonumber\\
 X^\dagger &= & x_0 -x_i\tau_i.
\label{prconj}\end{eqnarray} On the seven sphere, infinitesimal
homogeneous transformations which play the role of the Lorentz
algebra can be introduced through
\begin{eqnarray}
\delta X &=& a\cdot X,\end{eqnarray} with $a$ an infinitesimal
constant octonion. The requirement of preserving the unitary norm
(\ref{uninorm}) implies the vanishing of the $a_0$ component, so
that $a \equiv a_i\tau_i$.

\section{The octonionic $M$-superalgebra}

The octonionic $M$-superalgebra is introduced by assuming an
octonionic structure for the spinors which, in the $D=11$
Minkowskian spacetime, are octonionic-valued $4$-component
vectors. The algebra replacing (\ref{Malg}) is given by
\begin{eqnarray}
\{Q_a, Q_b\} = \{{Q^\ast}_a, {Q^\ast}_b\} =0,&\quad &\{Q_a,
{Q^\ast}_b\} = Z_{ ab},\label{divalgsusy}
\end{eqnarray}
where $\ast$ denotes the principal conjugation in the octonionic
division algebra and, as a result, the bosonic abelian algebra on
the r.h.s. is constrained to be hermitian
\begin{eqnarray}
Z_{ab}& =& {Z_{ba}}^\ast,
\end{eqnarray}
leaving only $52$ independent components.\par The
 $Z_{ab}$ matrix can be represented either as the $11+41$
bosonic generators entering
\begin{equation}\label{eq1}
  {{Z}}_{ab} = P^\mu (C\Gamma^{}_\mu)_{ab} +
   Z^{\mu\nu}_{\bf{O}} (C\Gamma^{}_{\mu\nu})_{ab}
   ,
\end{equation}
or as the $52$ bosonic generators entering
\begin{equation}\label{eq2}
  {{Z}}_{ab} =
    Z^{[\mu_1\ldots \mu_5]}_{\bf{O}}
    (C\Gamma^{}_{\mu_1 \ldots
    \mu_5})_{ab}\, .
\end{equation}
Due to the non-associativity of the octonions, unlike the real
case, the sectors individuated by (\ref{eq1}) and (\ref{eq2}) are
not independent. Furthermore, as we have already seen for $k=2$,
in the antisymmetric products of $k$ octonionic-valued matrices, a
certain number of them are redundant (for $k=2$, due to the $G_2$
automorphisms, $14$ such products have to be erased). In the
general case \cite{crt} a table can be produced expressing the
number of independent components in $D$ odd-dimensional spacetime
octonionic realizations of Clifford algebras, by taking into
account that out of the $D$ Gamma matrices, $7$ of them are
octonionic-valued, while the remaining $D-7$ are purely real. We
get the following table, with the columns labeled by $k$, the
number of antisymmetrized Gamma matrices and the rows by $D$ (up
to $D=13$) {\small {{\begin{eqnarray}&
\begin{tabular}{|c|c|c|c|c|c|c|c|c|c|c|c|c|c|c|}\hline
$D\setminus k
$&$0$&$1$&$2$&$3$&$4$&$5$&$6$&$7$&$8$&$9$&$10$&$11$&$12$&$13$\\
\hline $7$&$1$&$7$&$7$&$1$&$1$&$7$&$7$&$1$&&&&&&\\ \hline

$9$&$1$&$9$&$22$&$22$&$10$&$10$&$22$&$22$&$9$&$1$&&&&\\ \hline

$11$&$1$&$11$&$41$&$75$&$76$&$52$&$52$&$76$&$75$&$41$&$11$&$1$&&\\
\hline

$13$&$1$&$13$&$64$&$168$&$267$&$279$&$232$&$232$&$279$&$267$&$168$&$64$&$13$&1\\
\hline
\end{tabular}&\nonumber\end{eqnarray}}} }
\begin{eqnarray}&&\end{eqnarray}
For what concerns the octonionic equivalence of the different
sectors, it can be symbolically expressed, in different odd
space-time dimensions, according to the table {{\begin{eqnarray}&
\begin{tabular}{|c|c|}\hline

$D=7$& $M0\equiv M3$\\ \hline

$D=9$&$M0+M1\equiv M4$\\ \hline

$D=11$&$M1+M2\equiv M5$\\ \hline

$D=13$&$M2+M3\equiv M6$\\ \hline

$D=15$&$M3+M4\equiv M0+M7$\\ \hline

\end{tabular}&\label{tablem}\end{eqnarray}}}

In $D=11$ dimensions the relation between $M1+M2$ and $M5$ can be
made explicit as follows. The $11$ vectorial indices $\mu$ are
split into the $4$ real indices, labeled by $a,b,c,\ldots$ and the
$7$ octonionic indices labeled by $i,j,k,\ldots$. The $52$
independent components are recovered from $52=4+2\times 7+6+28$,
according to {{\begin{eqnarray}&
\begin{tabular}{|c|c|c|}\hline

$4$ &$M1_a$ &$M5_{[aijkl]}\equiv {M5}_a$ \\ \hline

$7$&$M1_i$, $ M2_{[ij]}\equiv M2_{i}$    & $M5_{[abcdi]} \equiv
M5_i$,$ M5_{[ijklm]}\equiv {\widetilde M5}_{i}$
  \\ \hline

$6$&$M2_{[ab]}$& $M5_{[abijk]}\equiv M5_{[ab]}$   \\ \hline

$4\times 7= 28$&$M2_{[ai]}$& $M5_{[abcij]}\equiv M5_{[ai]}$
\\ \hline

\end{tabular}&\nonumber\end{eqnarray}}}
\begin{eqnarray}
&&
\end{eqnarray}

\section{The octonionic superconformal $M$-algebra}

The conformal algebra of the octonionic M-theory can be introduced
\cite{lt2} adapting to the eleven dimensions the procedure
discussed in \cite{cs} for the $10$ dimensional case. It requires
the identification of the conformal algebra of the octonionic
$D=11$ $M$-algebra with the generalized Lorentz algebra in the
$(11,2)$-dimensional space-time. In such a space-time the
octonionic Clifford's Gamma-matrices are $8$-dimensional. The
basis of the hermitian generators is given by the $64$
antisymmetric two-tensors
$
C\Gamma_{[\mu_1\mu_2]}{\cal Z}^{\mu_1\mu_2} $ and the $168$
antisymmetric three tensors $ C\Gamma_{[\mu_1\mu_2\mu_3]}{\cal
Z}^{\mu_1\mu_2\mu_3} $ (or, equivalently, by the $232$
antisymmetric six-tensors $ C\Gamma_{[\mu_1\ldots \mu_6]}{\cal
Z}^{\mu_1\ldots\mu_6} $). This is already an indication that the
total number of generators in the conformal algebra is $232$. We
will show that this is the case.

According to \cite{cs} the conformal algebra can be introduced as
the algebra of transformations leaving invariant the inner product
of Dirac's spinors. In $(11,2)$ this is given by $\psi^\dagger C
\eta$, where the matrix $C$, the analogous of the $\Gamma^0$,
given by the product of the two space-like Clifford's Gamma
matrices, is real-valued and totally antisymmetric. Therefore, the
conformal transformations are realized by the octonionic-valued
$8$-dimensional matrices ${\cal M}$ leaving $C$ invariant, i.e.
satisfying
\begin{eqnarray}
{\cal M}^\dagger C + C {\cal M} &=& 0.
\end{eqnarray}
This allows identifying the (quasi)-group of conformal
transformations with the (quasi-)group of symplectic
transformations. Indeed, under a simple change of variables, $C$
can be recast in the form
\begin{eqnarray}
\Omega &=&\left(
\begin{array}{cc}
0& {\bf 1}_4 \\ - {\bf 1}_4 & 0
\end{array} \right).
\end{eqnarray}
The most general octonionic-valued matrix leaving invariant
$\Omega$ can be expressed through
\begin{eqnarray}
{\bf M} &=&\left(
\begin{array}{cc}
D& B \\ C & -D^\dagger
\end{array} \right),\label{confM}
\end{eqnarray}
where the $4\times 4$ octonionic matrices $B$, $C$ are hermitian
\begin{eqnarray}
B=B^\dagger,\quad &&\quad C=C^\dagger .\label{BCcond}
\end{eqnarray}
It is easily seen that the total number of independent components
in (\ref{confM}) is precisely $232$, as we expected from the
previous considerations.

It is worth noticing that the set of matrices ${\bf M}$ of
(\ref{confM}) type forms a closed algebraic structure under the
usual matrix commutation. Indeed $\relax [ {\bf M}, {\bf M}]
\subset {\bf M}$ endows the structure of $Sp(8|{\bf O})$ to ${\bf
M}$. For what concerns the supersymmetric extension of the
superconformal algebra, we have to accommodate the $64$ real
components (or $8$ octonionic) spinors of $(11,2)$ into a
supermatrix enlarging $Sp(8|{\bf O})$. This can be achieved as
follows. The two $4$-column octonionic spinors $\alpha$ and
$\beta$ can be accommodated into a supermatrix of the form
\begin{eqnarray}&&
\left(\begin{array}{c|cc}
  0 & -\beta^\dagger & \alpha^\dagger\\ \hline
  \alpha & 0& 0 \\
  \beta & 0 & 0
\end{array}\right)\label{fermionic}.
\end{eqnarray}
Under anticommutation, the lower bosonic diagonal block reduces to
$Sp(8|{\bf O})$. On the other hand, extra seven generators,
associated to the $1$-dimensional antihermitian matrix $A$
\begin{eqnarray}
A^\dagger &=& - A, \label{Acond}
\end{eqnarray}
i.e. representing the seven imaginary octonions, are obtained in
the upper bosonic diagonal block. Therefore, the generic bosonic
element is of the form
\begin{eqnarray}&&
\left(\begin{array}{c|cc}
  A & 0 & 0\\ \hline
  0 & D& B \\
  0& C & -D^\dagger
\end{array}\right)\label{bosonic},
\end{eqnarray}
with $A$, $B$ and $C$ satisfying (\ref{Acond}) and
(\ref{BCcond}).\par The closed superalgebraic structure, with
(\ref{fermionic}) as generic fermionic element and (\ref{bosonic})
as generic bosonic element, will be denoted as $OSp(1,8|{\bf O})$.
It is the superconformal algebra of the $M$-theory and admits a
total number of $239$ bosonic generators.

\section{Conclusions.}

We have seen that, contrary to what is commonly believed, an
alternative formulation for the $M$ superalgebra and the $M$
superconformal algebra can be consistently introduced in
association with the non-associative maximal division algebra of
the octonions. It presents peculiar features, like the
non-independence of the different octonionic brane sectors, which
is a reflection of the higher-rank antisymmetric octonionic
tensorial identities discussed in section {\bf 3}. The existence
of this second variant of the $M$ algebra is puzzling. It could be
ultimately related with the arising of exceptional structures
(exceptional Lie and Jordan algebras) in the ``Theory Of
Everything" \cite{boy}.\par Since imaginary octonions admits a
geometrical description in terms of the seven sphere $S^7$, it
could be speculated that the higher-dimensional octonionic
descriptions, e.g. of the eleven dimensions, corresponds to a
particular compactification of the eleven-dimensional $M$ theory
down to $AdS_4\times S^7$. This compactification corresponds to a
natural solution for the $11$ dimensional supergravity, see
\cite{dp}.
\par The octonionic superconformal algebra $OSp(1,8|{\bf O})$ has
been explicitly derived. It corresponds to a supersymmetric
extension of a bosonic conformal algebra which is mathematically
interesting since it corresponds to a closed algebraic structure
which goes beyond the standard notion of conformal algebra of a
given Jordan algebra, see \cite{lt2}.


\begin{thebibliography}{0}

\bibitem{hls} R. Haag, J. \L opusza\'{n}ski and M. Sohnius,{\it
Nucl.Phys.} {\bf B 88}, (1975), 257.
\bibitem{kt} T. Kugo and P. Townsend, {\it Nucl. Phys.} {\bf B 221}
(1983), 357.
\bibitem{fer} R. D'Auria, S. Ferrara, M.A. Lledo and V.S. Varadarajan,
J. Geom. Phys. {\bf 40} 101 (2001); R. D'Auria, S. Ferrara and
M.A. Lledo, Lett. Math. Phys. {\bf 57}, 123 (2001); S. Ferrara and
M.A. Lledo, Rev. Math. Phys. {\bf 14}, 519 (2002).
\bibitem{fm} D.B. Fairlie and A.C. Manogue, {\it Phys. Rev.} {\bf D 34},
1(1986), 1832.
\bibitem{cs} K.W. Chung and A. Sudbery, {\it Phys. Lett.} {\bf B 198},
(1987), 161.
\bibitem{bs} C.A. Barton and A. Sudbery, math.RA/0203010.
\bibitem{wit} E. Witten, ``Quest For Unification", hep-ph/0207124.
\bibitem{ram} P. Ramond, ``Algebraic Dreams", hep-th/0112261.
\bibitem{cmp} M. G\"{u}naydin, C. Piron and H. Ruegg, {\it Comm.
Math. Phys.} {\bf 61} (1978), 69.
\bibitem{smo} L. Smolin, hep-th/0104050.
\bibitem{lt1} J. Lukierski and F. Toppan, {\it Phys. Lett.} {\bf B 539},
(2002), 266.
\bibitem{lt2} J. Lukierski and F. Toppan, ``Octonionic $M$-theory
and $D=11$ Generalized Conformal and Superconformal Algebras",
hep-th/0212201.
\bibitem{bst} E. Bergshoeff, E. Sezgin, P, Townsend, {\it Ann.
Phys.} {\bf 18}, 330 (1987).
\bibitem{dk} M.J. Duff. R.R. Khuri, J.X. Lu, {\it Phys. Rep.} {\bf
259}, 213 (1995).
\bibitem{gk} M. G\"{u}naydin and S.V. Ketov, {\it Nucl. Phys.} {\bf B
467} (1996), 215.
\bibitem{hl} J.Lukierski and P. Minnaert, {\it Phys. Lett.} {\bf B 129}
(1983), 392; Z. Hasiewicz and J. Lukierski, {\it Phys. Lett.} {\bf
B 145} (1984), 65.
\bibitem{cp}  M. Cederwall and C. Preitschopf, {\it Comm. Math. Phys.} {\bf
167} (1995), 373.
\bibitem{crt} H.L. Carrion, M. Rojas and F. Toppan, ``Quaternionic
and Octonionic Spinors. A classification", hep-th/0302113.
\bibitem{boy} L. Boya, ``Octonions and M-theory", hep-th/0301037.
\bibitem{dp} M.J. Duff and C.N. Pope, ``Kaluza-Klein supergravity
and the seven-sphere" in Supersymmetry and Supergravity 82,
Trieste proceedings; M.J. Duff, B.E.W. Nilsson and C.N. Pope, {\it
Phys. Rep.} {\bf 130} (1986), 1.
\end{thebibliography}
\end{document}